\documentclass[twocolumn,showpacs,preprintnumbers,amsmath,amssymb,
showkeys,superscriptaddress]{revtex4}
\usepackage{graphicx}
\usepackage{dcolumn}
\usepackage{bm}
\usepackage{bigints}
\usepackage{natbib}
\usepackage{relsize}
\usepackage{amsmath}
\usepackage{placeins} 

\usepackage{xcolor}
\usepackage{hyperref}

\begin{document}

\title{Prospects for an independent measurement of the $\boldsymbol{\ell} = 1, 2, 3$ CMB anisotropy multipoles using the anisotropic Sunyaev-Zel'dovich effect}

\author{D.I. Novikov}
\affiliation{Astro-Space Center of P.N. Lebedev Physical Institute, Profsoyusnaya 84/32, Moscow 117997, Russia}

\author{T.I. Larchenkova}
\affiliation{Astro-Space Center of P.N. Lebedev Physical Institute, Profsoyusnaya 84/32, Moscow 117997, Russia}

\author{A.O. Mihalchenko}
\affiliation{Astro-Space Center of P.N. Lebedev Physical Institute, Profsoyusnaya 84/32, Moscow 117997, Russia}

\author{A.M. Osipova}
\affiliation{Astro-Space Center of P.N. Lebedev Physical Institute, Profsoyusnaya 84/32, Moscow 117997, Russia}
\affiliation{HSE University, 101000 Moscow, Russia}

\author{K.O. Parfenov}
\affiliation{Astro-Space Center of P.N. Lebedev Physical Institute, Profsoyusnaya 84/32, Moscow 117997, Russia}

\author{S.V. Pilipenko}
\affiliation{Astro-Space Center of P.N. Lebedev Physical Institute, Profsoyusnaya 84/32, Moscow 117997, Russia}
\affiliation{HSE University, 101000 Moscow, Russia}


\begin{abstract}
  We investigate the prospects for observing a specific spectral distortion
  of the cosmic microwave background, which occurs due to the anisotropy of
  the radiation when it is scattered by
  hot plasma of galaxy clusters.
  Detection of this anisotropic Sunyaev-Zel'dovich effect would enable
  an independent measurement of the CMB anisotropy multipoles with
  $\ell = 1, 2, 3$, allow the Sachs-Wolfe and integrated Sachs-Wolfe
  (Rees-Sciama) effects to be separated, and partially solve the cosmic variance problem for low multipoles. We applied the Least Response Method for component
  separation in the data processing and estimated the required sensitivity
  of the experiment for such observations. We test our approach on a
  simulated signal that is contaminated by various foregrounds with poorly defined spectral shapes, along with distortions of the relic blackbody spectrum caused by the Sunyaev-Zel'dovich effect and its relativistic corrections. 
\end{abstract}


\maketitle

\section{Introduction}Measurements of the cosmic microwave background (CMB) temperature anisotropy by WMAP \citep{2013ApJS..208...20B} and Planck \citep{2014A&A...571A..15P,Planck_2018_4} show 
alignment of the quadrupole and octupole \citep{2004PhRvD..70d3515C,2006MNRAS.367...79C,2008IJMPD..17..179N} and a lack of power of
multipoles with low $\ell$ in the observed angular spectrum
$C_{\ell}$ \citep{2003MNRAS.346L..26E,2003PhRvD..68l3523T,2004PhRvL..93v1301S,2014A&A...571A..15P}. In addition, according to the studies \citep{2010AdAst2010E..92C, 2014PhRvD..90j3510N, 2016NuPhB.906..367W, 2016A&A...594A..16P, 2016CQGra..33r4001S, 2021JCAP...03..103C}, there are
other large-scale anomalies in the temperature maps of the relic
radiation. These observations are difficult to reconcile with the predictions of Gaussian statistics and the standard inflationary model.
Such anomalies may be due to incomplete removal of foreground components, particularly those from the Milky Way. The intrinsic cosmological dipole is completely masked by the dipole induced by our motion with respect to the CMB rest frame. Besides, the observed anisotropy map represents only one realization of a random process, which inevitably leads to the ``cosmic variance'' problem for low multipoles.

This motivates the search for an independent source of information on the low-order anisotropy multipoles beyond the directly observed CMB sky. This source of information should not be so sensitive to the specific spatial distribution of foregrounds in our vicinity, should not
depend on our motion relative to the CMB rest frame, and should help us at least partially get around the cosmic variance problem.

One possible way to obtain such alternative information on low multipoles
is to observe the anisotropic Sunyaev-Zel'dovich effect (aSZ) described in
\citep{2018PhRvD..98l3513E, 2020PhRvD.101l3510N}. The impact of radiation anisotropy on the scattered frequency spectrum is also investigated in \citep{2012MNRAS.426..510C}. This effect is a
correction to the well-known thermal Sunyaev-Zel'dovich effect (tSZ) \citep{1969Ap&SS...4..301Z,1970Ap&SS...7...20S} due to
the anisotropy of radiation scattered by hot electron plasma of galaxy clusters.

According to \citep{1969JPhys..30..301B}, the first correction to the Thomson scattering cross section is
proportional to the electron temperature $\Theta_e=k_{_{B}}T_e/m_ec^2$ and effectively
depends on the scattering
angle. Using this fact, it is shown in \citep{2018PhRvD..98l3513E, 2020PhRvD.101l3510N} that in the linear approximation in
$\Theta_e$ the dipole, quadrupole, and
octupole components of the anisotropic radiation incident on the cluster
create a distinctive spectral distortion
of the scattered radiation. The amplitude of this distortion is
proportional to $\sim\Theta_e\cdot\Delta T/T$, where $\Delta T/T$ is the
anisotropy of the radiation at the scattering point. Thus, the
isotropic monopole fraction of the radiation is responsible for the classical
thermal effect, while the anisotropy of the radiation creates an additional
anisotropic effect of a different spectral shape.

For an observer located in nearby clusters, the
anisotropy map of the CMB should be nearly identical to that observed
at our location including
the same  $\Delta T/T$ multipoles in both amplitude and orientation with
$\ell=1,2,3$. Thus, observation of the CMB radiation scattered by nearby
clusters with distorted frequency spectrum gives us a unique opportunity of an independent
measurement of the CMB dipole, quadrupole, and octupole.

As for distant clusters, according to \cite{2018PhRvD..98l3513E, 2020PhRvD.101l3510N}, observing such a distortion of the
spectrum on a large number of clusters is a way to separate the Sachs-Wolfe
effect (SW) \cite{1967ApJ...147...73S} from the Integrated Sachs-Wolfe effect (ISW) \cite{1968Natur.217..511R} and to some extent
get rid of
the cosmic variance for multipoles with $\ell=1,2,3$. This is explained
by the fact that the integrated effect is formed
at very low redshifts and, thus, the spectral distortions
caused by the anisotropic SZ effect from high redshift SZ
clusters are ISW free.

In our article, we evaluate the practical possibility of detecting an aSZ
signal in a real experiment. In this case, the multicomponent nature of
the observed signal must be considered, accounting for both relativistic corrections to the SZ effect and foreground signals of various origins.

Since the 1990s, numerous studies have been published addressing various corrections to the Kompaneets equation
\citep{1957JETP...4.730K} and SZ effect. Several relativistic corrections
to the thermal SZ effect and multiple scattering
on SZ clusters are considered in \citep{2012MNRAS.425.1129C,2014MNRAS.437...67C,2014MNRAS.438.1324C,1998ApJ...499....1C,1998ApJ...502....7I,Stebbins:1997qr,Rephaeli:2002zs,2000astro.ph..5390I,1994PhRvD..49..648H}. A study of corrections for the kinematic SZ effect is carried out in   
\citep{1998ApJ...508...17N,1998ApJ...508....1S,1999ApJ...510..930C}.
An analytical investigation
of the Boltzmann equations was made and expressions for the photon redistribution functions are obtained in \citep{2009PhRvD..79h3005N,2013MNRAS.434..710N,2014MNRAS.441.3018N}. The additional correction to the SZ signal
due to the motion of the Solar System is discussed in
\citep{2005A&A...434..811C}. The influence of the CMB anisotropy on the spectral distortions is investigated in \citep{2016PhRvD..94b3513Y,Balashev:2015lla,2017PhRvL.119v1102Y, 2021PhRvL.127j1301F}.

These studies demonstrate that detecting the aSZ is complicated because it must be disentangled from other Sunyaev-Zel'dovich-related spectral distortions. Moreover, foreground signals,
including Galactic dust emission, the cosmic infrared background (CIB), synchrotron radiation, free-free emission, and potentially other sources, exceed
the target signal by several orders of magnitude, significantly
complicating observations and data analysis.

We consider a Fourier transform spectrometer as a target instrument in our modeling. Instruments of this kind have been proven to be suitable for the study of CMB spectral distortions \cite{1996ApJ...473..576F,2012A&A...538A..86D}. To minimize foregrounds, the instrument should be placed in space
\cite{2021PhyU...64..386N,2016SPIE.9904E..0WK,doi:10.1098/rsta.2020.0212}. 

Given the experiment's limited sensitivity, a careful
approach to data processing is especially important. Not only can this
approach facilitate detection of the aSZ signal, but it can also enable separation of other components, revealing the diverse effects arising during the
interaction of the CMB with the hot plasma of galaxy clusters. In addition,
employing optimal data processing methods for PIXIE-like experiments
\citep{2015PhRvL.115z1301H, 2016SPIE.9904E..0WK} 
enhances the detectability of $\mu$
distortions \citep{1970Ap&SS...9..368S, 1993PhRvL..70.2661H,2012JCAP...09..016K, 2012ApJ...758...76C, PhysRevD.111.043522, 2024JCAP...05..070M} originating
in the prerecombination epoch.

The data processing approach must account for our poor knowledge of the
expected spectra of some components (for example,
dust emission) whose parameters vary unpredictably
along the line of sight.

In \citep{2023PhRvD.107f3506N,2024PhRvD.109b3523M} the Least Response Method
(LRM) algorithm is developed for separating signals of interest from photon noise and foreground components with poorly defined spectra.
Other known methods such as Internal Linear Combination (ILC)
\citep{1992ApJ...398..169R} and its modifications: constrained ILC (cILC) and moment ILC (mILC) \citep{2011MNRAS.418..467R, 2020MNRAS.494.5734R, 2005MNRAS.357..145S, 10.1093/mnras/stx1982, 2021MNRAS.500..976R} have proven to be insufficiently effective for our
analysis. They either fail to remove foregrounds (ILC), or,
having removed them all (mILC), fail to sufficiently suppress the contribution of photon noise.
LRM minimizes the response
to all foregrounds and noise simultaneously, leaving the response to the
signal of interest constant. Unlike cILC or mILC, LRM does not require complete orthogonalization of the
signal of interest to all other components; that is, the linear filtering of the observed signal across
frequency channels does not completely cancel out the contribution from these components.
This property mitigates the impact of photon noise. As a result, the desired signal can be detected with
lower experimental sensitivity.

In our paper, we apply the LRM for finding the aSZ signal in simulated
data. The aim of this approach is to roughly estimate the required experimental sensitivity for aSZ observations and to test the mentioned multifrequency data processing method. In this initial analysis, we do not take into account
the kinematic SZ effect and its relativistic corrections.

The paper is organized as follows. In Sec. II, we review the
anisotropic Sunyaev-Zel'dovich effect and describe foreground models.
In Sec. III, we review the LRM method. In Sec. IV, we present our numerical results. Finally, in
Sec. V, we provide our brief conclusions.

\section{Anisotropic thermal SZ effect and foregrounds}
In this Section, we characterize (1) the spectral distortion generated by anisotropic radiation scattering off galaxy clusters as our signal of interest, and (2) the primary foreground components that challenge its detection.
\subsection{ASZ}
To describe the anisotropic SZ effect, we use the usual notations.
$I=I(\nu)$ is the spectral radiance, which is related to the photon number
density as $n(\nu)=\frac{c^2I}{2h\nu^3}$, where $\nu, c, h$  are the
frequency, the speed of light and the Planck constant respectively.
$T_e$ is the temperature of electrons, $T_r$ is the mean temperature of radiation,
$m_e$ is the rest mass of the electron, $N_e$ is the concentration
of electrons, $\sigma_{T}$ is the Thomson cross section and $\Theta_e=\frac{k_{_{B}}T_e}{m_ec^2}$,
where $k_{_{B}}$ is the Boltzmann constant.

We consider the scattering of anisotropic radiation on a stationary cloud
of hot electrons located within a galaxy cluster. The frequency spectrum
of the radiation before scattering
$n_0(\nu)$ has a blackbody shape:
\begin{equation}
  \begin{array}{l}
    \vspace{0.3cm}
    n_0(x)=B(x)=\frac{1}{e^x-1},\\
    x=\frac{h\nu}{k{_{_{B}}}T_r}.
  \end{array}
\end{equation}
We choose a local spherical coordinate system
$\boldsymbol{\Omega}(\theta,\varphi)$
with the center at the scattering point and the north pole pointing to the
observer. In this coordinate system, the temperature of the radiation
$T(\boldsymbol{\Omega})$ incident on the cluster can be decomposed into spherical harmonics $Y_{\ell m}$:
\begin{equation}
  \begin{array}{l}
 \frac{{T(\boldsymbol{\Omega})}-T_r}{T_r}=
    \sum\limits_{\ell m}
    \tilde{a}_{\ell m}\tilde{Y}_{\ell m}(\boldsymbol{\Omega}).
 \end{array}
\end{equation} 
Note that $\tilde{a}_{\ell m},\tilde{Y}_{\ell m}$ are not
the same as $a_{\ell m},Y_{\ell m}$ that belong to the
conventional Galactic coordinate system at our location.

In \citep{2018PhRvD..98l3513E}, it is shown that as a result of scattering
in the optically thin limit and for $T_r\ll T_e$,
$\Theta_e\ll 1$ the frequency spectrum distortion
$\Delta_n=n(x)-n_0(x)$ looks as follows:
\begin{equation}
  \begin{array}{l}

    \vspace{0.5cm}
    
    \Delta_n/\tau=\\

    \vspace{0.5cm}
    
    =(\gamma_1+\Theta_e\gamma_2)
    \hspace{0.1cm}\left(-x
    \hspace{0.1cm}\frac{dB}{dx}\right)+\hspace{1.2cm}\Big\}1\\

    \vspace{0.5cm}
        
    +\Theta_e\gamma_3\hspace{0.1cm}\frac{1}{x^2}\hspace{0.1cm}
    \frac{d}{dx}\left[x^4\frac{d}{dx}
      \left(-x\frac{dB}{dx}\right)\right]+\hspace{0.5cm}\Big\}2
    \hspace{0.3cm}(aSZ)\\

    \vspace{0.5cm}
    
    +\Theta_e\hspace{0.1cm}\frac{1}{x^2}\hspace{0.1cm}\frac{d}{dx}
    \left[x^4\frac{dB}{dx}\right],\hspace{2.3cm}\Big\}3\hspace{0.3cm}(tSZ)\\
    \tau=\int N_e\sigma_Tdr.
    
  \end{array}
\end{equation}
The third term in Eq. (3) represents the classical tSZ for the
isotropic fraction of radiation with the temperature
$T_r=\frac{1}{4\pi}\int T(\boldsymbol{\Omega})d\boldsymbol{\Omega}$,
which corresponds to the monopole $\ell=0$ in Eq. (2), while the first
two terms are due to the presence of CMB
anisotropy. According to \citep{2018PhRvD..98l3513E} only multipoles with $\ell=1,2,3$ distort
the signal.  The coefficients $\gamma_1,\gamma_2,\gamma_3$
depend on the magnitude and orientation of the CMB dipole, quadrupole and
octupole at the position of scattering:
\begin{equation}
  \begin{array}{l}
    \vspace{0.3cm}
    
    \gamma_1=\frac{1}{4\sqrt{\pi}}\left(\frac{1}{\sqrt{5}}\tilde{a}_{20}
     -2\frac{\Delta_T(\boldsymbol{\Omega_z})}{T_r}\right),\\

    \vspace{0.3cm}
    
    \gamma_2=-\frac{3}{2\sqrt{\pi}}\left(
    \frac{2}{5\sqrt{3}}\tilde{a}_{10}+\frac{1}{\sqrt{5}}\tilde{a}_{20}
    -\frac{2}{5\sqrt{7}}\tilde{a}_{30}\right),\\
    
   \gamma_3=-\frac{3}{4\sqrt{\pi}}\left(\frac{4}{5\sqrt{3}}\tilde{a}_{10}-
    \frac{1}{3\sqrt{5}}\tilde{a}_{20}+
    \frac{1}{5\sqrt{7}}\tilde{a}_{30}\right),
  \end{array}
\end{equation}
where $\boldsymbol{\Omega_z}$ is the direction along the z-axis to the observer
(to the north pole) and $\Delta_T(\boldsymbol{\Omega})=T(\boldsymbol{\Omega})-T_r$. It should be
noted that the spectral distortions are produced only by the anisotropy of
the incident radiation that is symmetrical relative to the z axis, i.e.
by components with m=0.

The deviation of the form $-x\frac{dB}{dx}$ [the first term in Eq. (3)]
is of no interest to us since its shape coincides with the spectral
variations caused by the observed temperature anisotropy. Indeed
$\frac{dB}{dT}=-\frac{x}{T}\frac{dB}{dx}$. At the same time, the distortion
$\sim\frac{1}{x^2}\hspace{0.1cm}\frac{d}{dx}\left[x^4\frac{d}{dx}
  \left(-x\frac{dB}{dx}\right)\right]$ [the second term in Eq. (3)] has a
very characteristic spectral form, which allows us to separate it from
other components of the observed signal. For convenience, in \citep{2018PhRvD..98l3513E},
this distortion is called
the aSZ.  It is important to note that
according to Eq. (3), the amplitudes of the anisotropic and thermal
effects are both proportional to $\tau\Theta_e$. That is, the ratio of these
amplitudes does not depend on the properties of the medium of the observed
galaxy cluster. This makes it possible to directly estimate the
coefficient $\gamma_3$ when both effects are observed simultaneously.
Thus, the component of the spectrum observed from a cluster of galaxies in
the form of the aSZ is considered as the signal of interest:
\begin{equation}
  \begin{array}{l}
    \vspace{0.3cm}
    I_{asz}(\nu)=A_{asz}\hspace{0.1cm}x\frac{d}{dx}\left[x^4\frac{d}{dx}
      \left(-x\frac{dB}{dx}\right)\right],\\
    A_{asz}=\frac{2(k_{_{B}}T_r)^3}{(hc)^2}\tau\Theta_e\gamma_3
 \end{array}
\end{equation}
Here the values toward a rich cluster of galaxies are as follows: $\tau \approx 1/100$,  $\Theta_e = 7.5\, \text{keV} / 511 \,\text{keV}$ and $\gamma_3 \sim 5 \cdot 10^{-6}$. Hence, $A_{asz} \approx 0.2$ Jy/sr.

Considering the fact that the dipole, quadrupole and octupole contribute to the amplitude of the same frequency
distortion, separating these contributions is only possible
by measuring the spectrum from several clusters close to us, i.e., from clusters where these multipoles have the
same fixed orientations and $\Delta T/T$ amplitudes as at our location.
 In total, it is necessary to measure such a distortion from at least
 15 clusters in order to separate the contributions from $a_{\ell m}$, $\ell=1,2,3$, $-\ell\le m\le\ell$
 (from three dipole components, five quadrupole components and seven octupole components).
 Using the expression for $\gamma_3$ in Eq. (4), and also taking into account that
 $a_{\ell m}\sim1/\sqrt{\ell(\ell+1)}$ for low multipoles, one can estimate the magnitudes of the
 contributions from $\ell=1,2,3$ anisotropy to the overall effect:
\begin{equation}
\begin{array}{l}
  \vspace{0.2cm}
  A_{asz}= A_{asz}^{\ell=1}+A_{asz}^{\ell=2}+A_{asz}^{\ell=3},\\
  A_{asz}^{\ell=2}/A_{asz}^{\ell=1}\sim 0.19,\hspace{0.1cm}
  A_{asz}^{\ell=3}/A_{asz}^{\ell=1}\sim 0.07.
 \end{array}
\end{equation}
Therefore, the dipole component of the temperature anisotropy dominates this effect.
Detecting the contribution from the octupole component requires significantly greater
experimental sensitivity.
 \subsection{Foregrounds}

 In our approach, foregrounds are divided by their nature into two types:
 foregrounds with well-defined spectral shapes and foregrounds with poorly
 defined spectra. 
 Poorly defined spectra are parameterized spectra with
 floating parameters. However, these parameters can
 change only within a limited range of their possible variations. In
 addition, the amplitudes of all foregrounds cannot exceed certain
 preestimated values.
 \begin{figure*}[tbh]
  \includegraphics[width=0.425\textwidth]{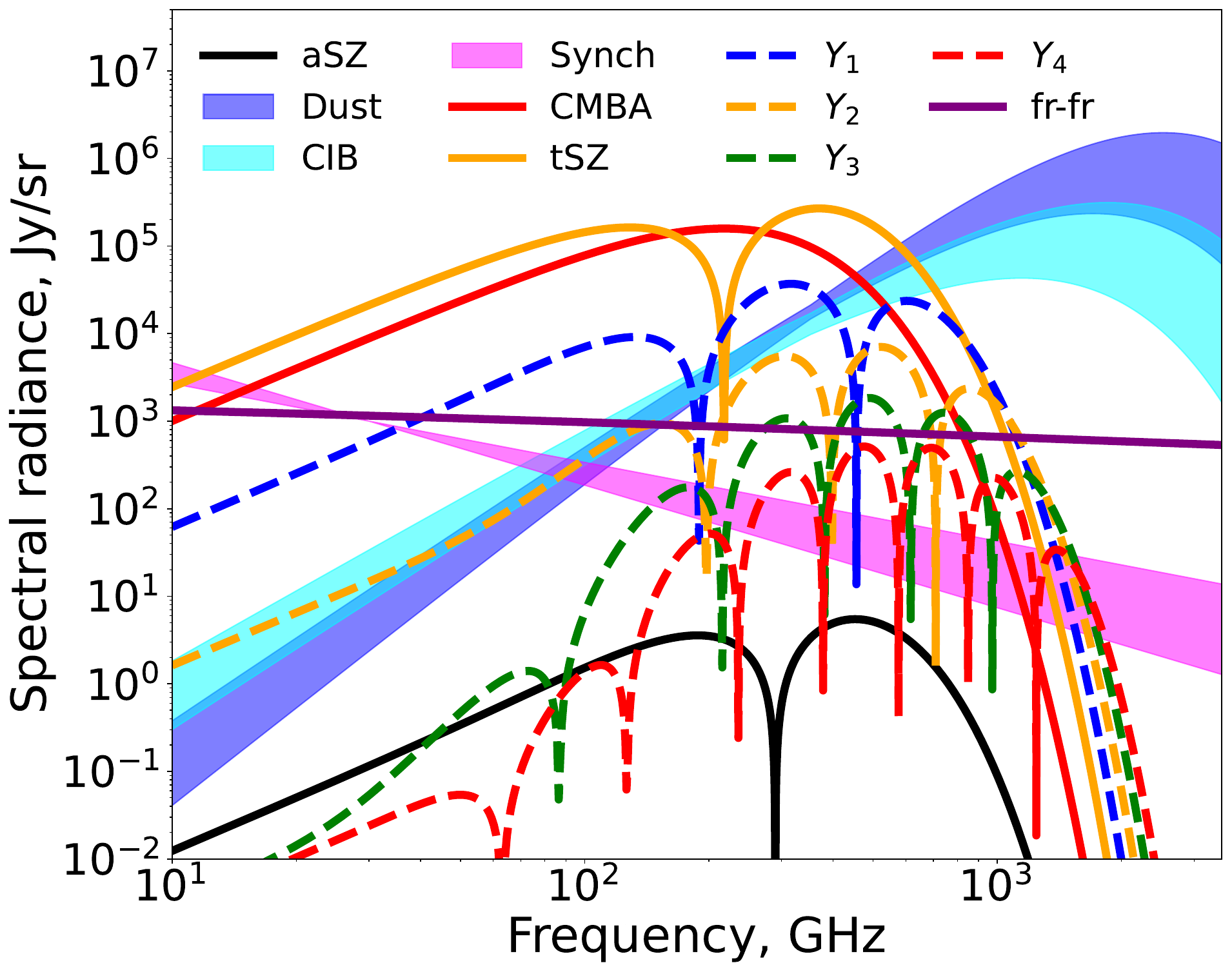}
  \includegraphics[width=0.49\textwidth]{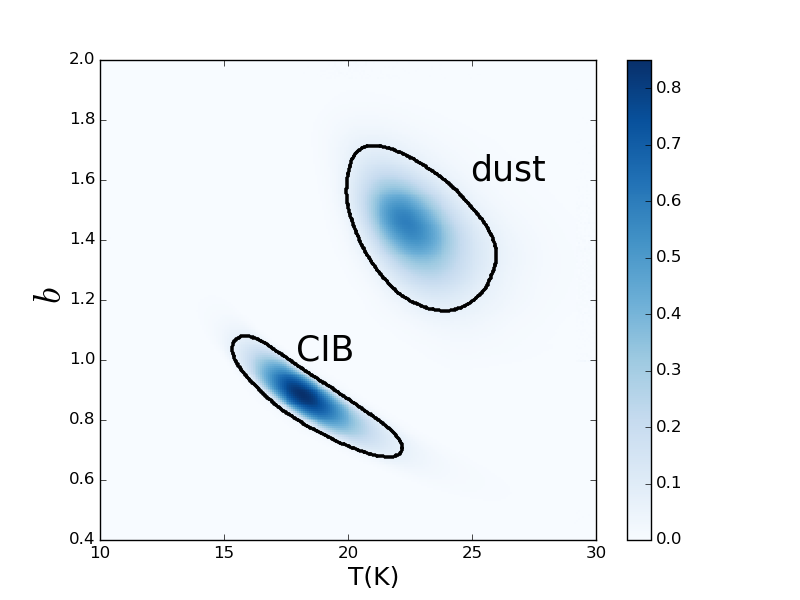}
  \caption{{\it Left panel}. Spectra of the aSZ signal and main foregrounds.
    Emissions from dust, the CIB and the synchrotron are poorly defined
    spectra with floating parameters. Shaded areas show possible
    variations of these spectra. The remaining spectra have strictly defined and well-known
    shapes: CMB anisotropy, thermal SZ effect, the first four relativistic corrections to it
     $Y_1,Y_2,Y_3,Y_4$  and the spectrum of free-free transitions.
    {\it Right panel}. Probability distribution function of the T and b parameters for dust and CIB at Galactic latitudes greater than $30^\circ$ or less than $-30^{\circ}$. The thermal dust is extracted from the Planck Data Release 3 (PDR3) "Commander" maps. The CIB parameters are constructed from PDR2 GNILC maps for CIB at 353, 545, and 857 GHz by fitting a three-parameter modified blackbody.
}
\end{figure*}
 Foregrounds with poorly defined spectra include dust emission, the cosmic
 infrared background (CIB) and synchrotron radiation.
 
 Dust and CIB are considered together and modeled as a
 modified black body radiation with two floating parameters: 
 temperature $T$ and spectral index $b$:
\begin{equation}
  \begin{array}{l}
    \vspace{0.2cm}
    I_{{}_{dust}}(\nu,T_d,b_d)=A_d(\nu/\nu_0)^{b_d}x_d^3 B(x_d),\hspace{0.1cm}
    x_d=\frac{h\nu}{k_{_{B}}T_d},\\
    I_{{}_{CIB}}(\nu,T_c,b_c)=A_c(\nu/\nu_0)^{b_c}x_c^3B(x_c),\hspace{0.1cm}
    x_c=\frac{h\nu}{k_{_{B}}T_c},\\
  \end{array}
\end{equation}
where $\nu_0=353$ GHz. The ranges of possible variations of
$T_d,b_d$  and $T_c,b_c$ are shown in Fig. 1 (right panel) for relatively
clean parts of the sky. 
The maximum allowable values for both $A_d$ and $A_c$ are estimated
as $3\cdot10^6$Jy/sr.

Synchrotron radiation is modeled according to
\citep{1986rpa..book.....R,delahoz23} and its spectrum is
\begin{equation}
  I_{{}_{sync}}(\nu,b_s)=A_s(\nu/\nu_s)^{-b_{s}},\\
\end{equation}
where $b_s$ is the only free parameter with possible variations
$0.9\le b_s\le 1.4$. Here $\nu_s=30$ GHz and $A_s \approx 1000$ Jy/sr. 
The values of $A_s$ and $b_s$ were extracted from QUIJOTE
\citep{delahoz23} sky maps at Galactic
latitudes greater than $30^\circ$ or less than $-30^{\circ}$.

For foregrounds with well-defined spectra, we consider free-free emission and
corrections associated with modifications of the blackbody spectrum of the
relic radiation.

Free-free emission is given by the following formula
\cite{2016A&A...594A..10P}:
\begin{equation}
  \begin{array}{l}
  I_{{}_{ff}}(\nu)=A_{ff}\left(1+\ln\left[1+\left(\frac{\nu_{ff}}{\nu}\right)
    ^{\sqrt{3}/\pi}\right]\right),\\
  \nu_{ff} = 255.33\;\mathrm{GHz} (T/10^3\;\mathrm{K})^{3/2},
  \end{array}
\end{equation}
where $A_{ff} =300$ Jy/sr and $T$ is the electron temperature. According to component maps
provided by the Planck Collaboration \cite{2016A&A...594A..10P}, $T$ is distributed between $6995$ 
and $7030$ K with a sharp peak at 7000 K. Consequently, we assume $\nu_{ff}\approx4728$ GHz and
this parameter does not vary much across the sky to noticeably change
the shape of the spectrum. Thus, in our calculations, we consider
free-free spectrum as a well-defined one.

Other foregrounds with well-defined spectra include the CMB temperature
anisotropy, which is the first derivative of $B(x)$ with respect to
radiation temperature $T_r$, and four relativistic corrections to the
thermal SZ effect:
\begin{equation}
   \begin{array}{l}
     \vspace{0.2cm}
      I_{{}_{CMBA}}(\nu)= A_{{}_{CMBA}}x^4\frac{dB}{dx},\\
      \vspace{0.2cm}
      I_{y_{0}}(\nu)=A_{y_0}x\frac{d}{dx}\left[x^4\frac{dB}{dx}\right],\\
    \vspace{0.2cm}
    I_{y_{n}}(\nu)=A_{y_n}\frac{x^4e^x}{(e^x-1)^2}Y_n(x),\hspace{0.5cm}n=1,..,4,\\
    A_{y_0}=\frac{2(k_{_{B}}T_r)^3}{(hc)^2}\tau\Theta_e,\hspace{0.3cm}A_{y_n}=
    A_{y_0}\Theta_e^n
    \vspace{0.2cm}
    \end{array}
\end{equation}
where $x=h\nu/k_{_{B}}T_r$, $T_r=2.72548$ $K$,
\citep{1990ApJ...354L..37M,Fixsen_2009}, $A_{CMBA}\le3.24\cdot10^4$ Jy/sr,
$A_{y_0} \approx 4\cdot 10^4$ Jy/sr.

The functions $Y_n(x)$, $n\ge 1$ for the relativistic
corrections have a rather cumbersome form and analytical formulas for them can be found in
\citep{1998ApJ...499....1C}.

Another additional term in the observed signal is photon noise, which appears
in each frequency channel as a random value. It comes from the CMB
radiation, a number of main foregrounds and emission from the instrument
including the optical system of the telescope. The magnitude of noise
and its possible correlations between different frequency channels can
be estimated in the process of observing the same galaxy cluster by
subtracting from each other the signals observed at different times over
the same time intervals.

The spectra of the foregrounds and the aSZ signal are presented in Fig. 1
(left panel).

\section{Component separation using the Least Response Method}

The Least Response Method is a linear filtering of the observed signal that minimizes the response to all foregrounds and noise while keeping the response to the desired signal constant.
It has the following distinctive features:\\
(1) It does not require complete orthogonalization of the various signal components. In other words, when
determining the contribution of any component, this method does not completely cancel out the contributions
of the remaining components with known spectra. This significantly reduces the impact of photon noise on
the recovered component amplitudes. For example, if we have two components, one of which is significantly larger than the other, when finding the larger component, the LRM practically ignores the smaller one, instead of setting it to zero and thereby increasing the response to white noise.\\
(2) It assumes our knowledge of the limits of possible variation of the foreground spectral parameters for those
foregrounds whose spectra are poorly defined.

As shown in the previous section, the observed frequency spectrum $S(\nu)$ can be divided
into two parts:
\begin{equation}
  \begin{array}{l}
    S(\nu)=I(\nu)+\tilde{I}(\nu),
 \end{array}   
\end{equation}
where $I(\nu)$ consists of components with well-defined spectral shapes,
while $\tilde{I}(\nu)$ includes components
with poorly defined spectra. In general, one can consider a total signal consisting of
$M$ components of different physical origin with well-defined spectra
and $K$ components with poorly defined spectra:
\begin{equation}
  \begin{array}{l}
    I(\nu)=\sum\limits_{m=1}^MI^m(\nu),\hspace{0.3cm}
    \tilde{I}(\nu)=\sum\limits_{k=1}^{K}\tilde{I}^k(\nu).
 \end{array}   
\end{equation}
Each component $I^m(\nu)$ with a fixed spectral shape is described either by a known analytical
function or by a strictly defined template $f^m(\nu)$:
\begin{equation}
  \begin{array}{l}
    I^m(\nu)=\alpha_mf^m(\nu),\hspace{0.3cm}m=1,...,M,
 \end{array}   
\end{equation}
where $\alpha_m$ is the amplitude of the $m$th component. Such components in the observed signal
may include, for example, the thermal SZ effect, relativistic corrections to it or the anisotropic SZ effect.

When processing observational data using the LRM method, it is assumed that the amplitudes
$\alpha_m$ cannot exceed certain preestimated limits:
\begin{equation}
  \begin{array}{l}
  \mid \alpha_m\mid \le A_m\hspace{0.3cm}m=1,...,M.\\
   \end{array}   
\end{equation}
Such upper limits on the amplitudes of various effects can be found from previous observations and
numerical estimates.

As for components with poorly defined spectra, they can usually also be described by analytical functions
$\tilde{f}^k(\nu,\boldsymbol{P_k})$, $k=1,..,K$. 
Unlike well-defined components, the shapes of their spectra depend on the set of parameters
$\boldsymbol{P_k}$ and change when these parameters are varied.
One example of such a component is the emission of thermal dust. The uncertainty in the spectra
arises from the integration of the signal along the line of sight, where the radiation from dust changes its
parameters, such as temperature or the slope of the spectrum.

Similar to Eq. (13), the observed spectrum from such components can be written as follows:
\begin{equation}
  \begin{array}{l}
    \vspace{0.2cm}
    \tilde{I}^{k}(\nu)=\frac{1}{V_{\tilde{\Omega}_k}}\int\limits_{\tilde{\Omega}_k}
    \tilde{\alpha}_{k}({\bf P_k})
    \tilde{f}^k(\nu,{\bf P_k})d{\bf P_k},\\
     \vspace{0.3cm}
    d{\bf P_k}=dp_{{}_1}^kdp_{{}_2}^k
    \cdot\cdot dp_{L_{k}}^k,\hspace{0.3cm}k=1,...,K,\\
    V_{{\tilde{\Omega}}_k}=\int\limits_{{\tilde{\Omega}}_k}d{\bf P_k},
 \end{array}   
\end{equation}
where ${\bf P_k}=p_{{}_1}^k,..,p_{L_k}^k$ is the set of $L_k$ parameters of the
$k$-th component and $\tilde{\Omega}_k$ is the range of possible variations
of these parameters. $\tilde{\Omega}_k$ for each of the poorly defined
components can be determined from the results of the previous
experiments such as Planck and WMAP. Analogously to Eq. (14) we also assume that the amplitudes of
the spectra $\tilde{\alpha}_k$ are less than some preestimated values:
\begin{equation}
  \begin{array}{l}
    \vspace{0.2cm}
   \mid \tilde{\alpha}_k({\bf P_k})\mid \le \tilde{A}_k
   \hspace{0.3cm}for\hspace{0.3cm}{\bf P_k}
   \in \tilde{\Omega}_k,
   \end{array}   
\end{equation}
and $\tilde{\alpha}_k({\bf P_k})=0$ otherwise. Therefore, we do not know the exact shapes of the
frequency spectra $\tilde{I}_k$ that contribute to the overall observed spectrum, but only the limitations
determined by Eqs. (15) and  (16) on the signal they create.

During the observations in  each sky pixel we will observe a discrete signal $S_j$ in $J$ frequency
channels $\nu_j$, consisting of the terms
\begin{equation}
  \begin{array}{l}
    \vspace{0.2cm}
    S_j=\sum\limits_{m=1}^M\alpha_mf_j^m+
    \sum\limits_{k=1}^K\tilde{I}_j^k+
    N_j,\hspace{0.2cm}j=1,..,J,
   \end{array}
\end{equation}
where $f_j^m=f^m(\nu_j)$, $\tilde{I}_j^k=\tilde{I}^k(\nu_j)$ and $N_j$ is the
photon noise. For convenience, this equation can be written in vector form:
\begin{equation}
  \begin{array}{l}
    \vspace{0.2cm}
  {\bf S}={\bf F}+{\bf\tilde{F}}+{\bf {N}},\\
 {\bf F}=\sum\limits_{m=1}^M\alpha_m{\bf f^{m}},
  \hspace{0.3cm}{\bf\tilde{F}}=\sum\limits_{k=1}^K{\bf\tilde{I}^k},
   \end{array}
\end{equation}
where bold symbols ${\bf X}$ denote row vectors
$\boldsymbol{X}=(X_1\cdot\cdot\cdot X_J)$.  Here $\boldsymbol{F}$ is the sum of components
with well-defined spectra, $\boldsymbol{\tilde{F}}$ contains
all foregrounds with poorly defined spectra and $\boldsymbol{N}$ is the
noise with zero mean and covariance matrix
$\left[C_{ij}\right]={\bf C}=\langle{\bf N^T}{\bf N}\rangle$.

Our task is to estimate with the best accuracy all the coefficients $\alpha_m$
for components
with well-defined spectral shapes, leaving untouched the poorly defined part of
the observed spectrum. That is, we need to find the vector
$\boldsymbol{\alpha}=(\alpha_1\cdot\cdot\cdot\alpha_M)$.
For each component number m, the LRM method finds the optimal
row vector of weights $\boldsymbol{\omega^m}=(\omega_1^m\cdot\cdot\cdot\omega_J^m)$. This vector ensures
for linear filtering the unit response to this component:
$\boldsymbol{\omega^m}\boldsymbol{{f^m}^T}=1$ and a minimal response to all other components and photon noise. Thus,
we can introduce a matrix of the size $J\times M$ whose columns are represented by the components of the
row vectors $\boldsymbol{\omega^m}$:
\begin{equation}
  \hspace{-0.3cm}\bold{W}=\begin{pmatrix}
  \boldsymbol{{\omega^1}^T}&\hspace{-0.2cm}\boldsymbol{{\omega^2}^T}
  &\hspace{-0.2cm}\cdot&\cdot&\boldsymbol{{\omega^M}^T}\\
  \end{pmatrix}
  =\begin{pmatrix}
  \vspace{0.1cm}
  \omega_1^1 & \omega_1^2 & . & . & \omega_1^M\\
   \vspace{0.1cm}
   \omega_2^1 & \omega_2^2 & . & . & \omega_2^M\\
    \vspace{0.1cm}
  . &  . & . & . & .\\
  \omega_J^1 & \omega_J^2 & . & . & \omega_J^M\\
\end{pmatrix}
\end{equation}
By multiplying the observed signal vector ${\bf S}$ by the matrix ${\bf W}$, we obtain an estimate for the
amplitude vector $\boldsymbol{\alpha}$:
\begin{equation}
  \begin{array}{l}
   \vspace{0.2cm}
         \boldsymbol{\overline{\alpha}}= {\bf S W}=\boldsymbol{\alpha}+\boldsymbol{R},
   \end{array}
\end{equation}
where the components $R_m$ of the vector $\bold{R}$ are as follows:
\begin{equation}
  \begin{array}{l}
   \vspace{0.2cm}
   R_m=\left[\sum\limits_{m'\ne m}\alpha_{m'}
     {\bf f^{m'}}+{\bf\tilde{F}}+{\bf N}\right]\boldsymbol{{\omega^m}^T},\\
   m=1,..,M,
   \end{array}
\end{equation}
and the summation is performed over all $m'$ from 1 to M except $m'=m$.
In \citep{2024PhRvD.109b3523M},  it is shown that the best weights for each $m$th component that minimize the response $R_m$ are:
\begin{equation}
   \begin{array}{l}
  \boldsymbol{\omega^m}={\bf f^mD^{-1}}\cdot
  \left({\bf f^m}{\bf D^{-1}}{\bf (f^m)^T}\right)^{-1},\\
  {\bf D}={\bf\Phi^m}+{\bf\tilde{\Phi}}+{\bf C}.
    \end{array}
\end{equation}
Here $\boldsymbol{\Phi^m}=[\Phi_{ij}^m]$ and
$\boldsymbol{\tilde{\Phi}}=[\tilde{\Phi}_{ij}]$
are the $J\times J$ covariance matrices of the foregrounds with well-defined and poorly defined spectra respectively. They are calculated as follows:
\begin{equation}
  \begin{array}{l}
     \vspace{0.3cm}
     \Phi_{ij}^m=\sum\limits_{m'\ne m}A_{m'}^2q_{ij}^{m'},
     \hspace{0.5cm}q_{ij}^{m'}=f_i^{m'}f_j^{m'},\\
     \vspace{0.3cm}
     \tilde{\Phi}_{ij}=
     \sum\limits_{k=1}^K\tilde{A}_k^2\tilde{q}_{ij}^k,\\
     \tilde{q}_{ij}^k=\frac{1}{V_{{\tilde{\Omega}}_k}}\cdot
     \int\limits_{{\tilde{\Omega}}_k}
     \tilde{f}_i^k({\bf P_k})\tilde{f}_j^k({\bf P_k})d{\bf P_k}.
 \end{array}   
\end{equation}
As can be seen from Eq. (20, 21), for each component $m$ the difference between its estimated amplitude
$\overline{\alpha}_m$ and its true value $\alpha_m$ is determined by the response
$R_m$. This response consists of the response to foregrounds $R_m^{fgr}$
(which are considered to be all other components with $m'\ne m$) and the response
to photon noise $R_m^{Noise}$:
\begin{equation}
  \begin{array}{l}
    \vspace{0.2cm}
    \overline{\alpha}_m-\alpha_m=
    R_m=R_m^{fgr}+R_m^{Noise},\hspace{0.2cm}m=1,..,M,\\
     \vspace{0.2cm}
   R_m^{fgr}=\left[\sum\limits_{m'\ne m}\alpha_{m'}
     {\bf f^{m'}}+{\bf\tilde{F}}\right]\boldsymbol{{\omega^m}^T},\\
   R_m^{Noise}={\bf N}\boldsymbol{{\omega^m}^T}.\\
   \end{array}
\end{equation}
In approaches such as cILC and mILC, the  weights $\boldsymbol{\omega^m}$ are calculated with
the strict condition of zeroing out the contributions from all foregrounds
selected for exclusion from the signal: $R_m^{fgr}=0$. This leads
to an extremely uneven distribution of the weights $\omega_j^m$ and their large values.
This significantly increases the response to photon noise $R_m^{Noise}$. LRM, on the other hand, optimally distributes the response to foregrounds and noise in such a way that their combined response is
minimized.

The approach described above not only estimates the amplitudes of signal
components better than cILC, but also has the potential for improvement.
To find the amplitudes $\alpha_m$, we have safely used the upper possible
estimate for all other amplitudes $\alpha_{m'}$, $m'\ne m$. Refining the values
$\alpha_m$  can in principle improve the result.

\begin{figure*}[!htbp]
  \includegraphics[width=0.49\textwidth]{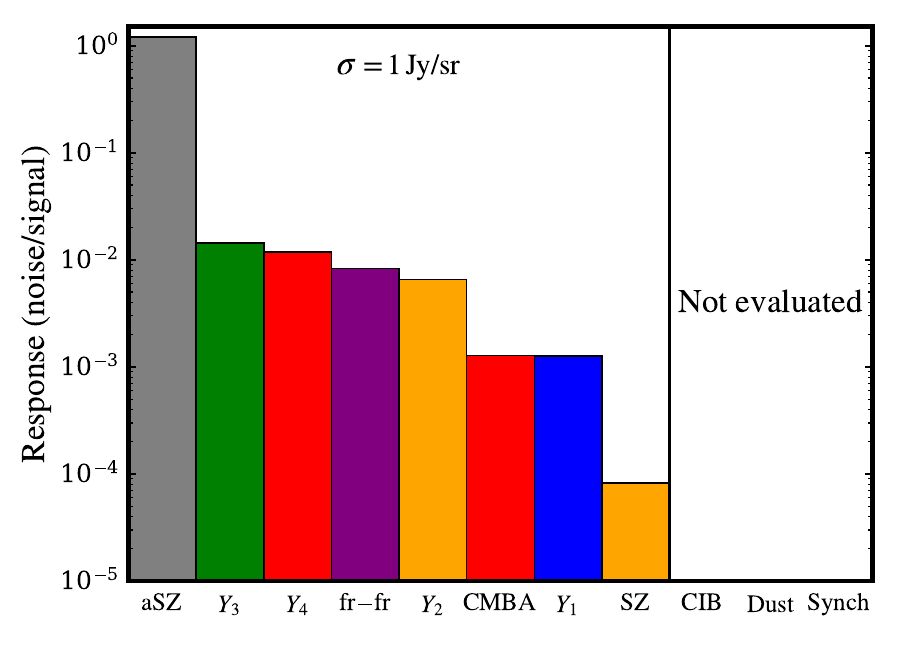}
  \includegraphics[width=0.49\textwidth]{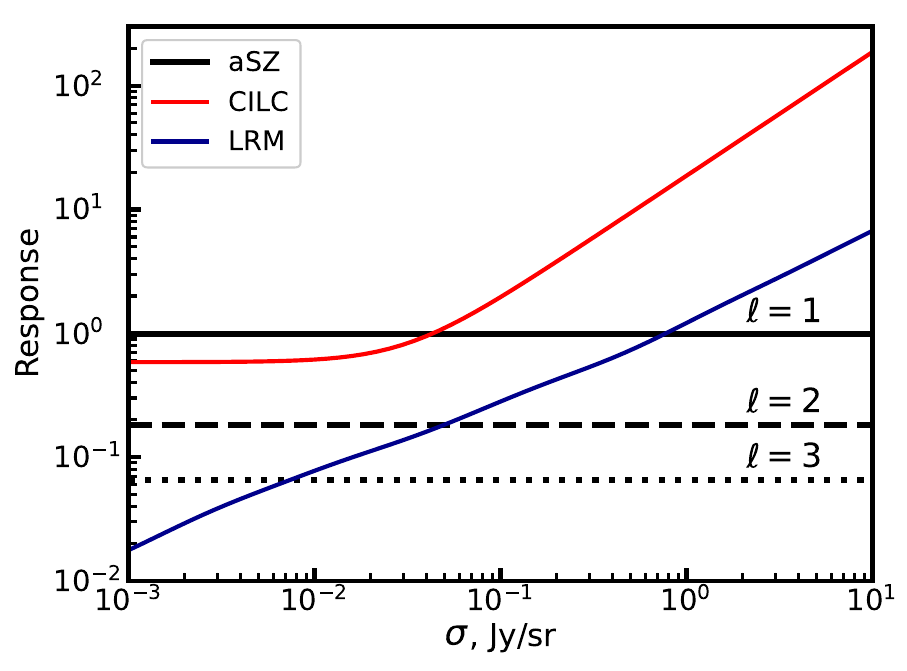}
  \caption{Application of the LRM method for signal component separation and
    extraction of the aSZ from the observed spectrum:\\
    {\it Left panel}. The histogram shows the total response to foregrounds and photon
    noise when extracting a signal with a well-defined spectral shape. The response
    is equivalent to the noise-to-signal ratio. Foregrounds with poorly defined
    spectral shapes were not extracted from the observed spectrum. The results were
    obtained for a sensitivity of 1 Jy/sr per frequency channel using the LRM method.\\
    {\it Right panel}. The red line shows the total response to all foregrounds and
    noise when extracting the aSZ signal using the cILC approach. The blue line
    corresponds to the same when using the LRM method. The black horizontal
    lines correspond to the aSZ signal caused by the dipole ($\ell=1$),
    quadrupole ($\ell=2$), and octupole ($\ell=3$) components of the
    CMB anisotropy at the location of the galaxy cluster. The intersection points
    with the horizontal lines correspond to a signal-to-noise ratio of 1.}
\end{figure*}

\section{Numerical results}

For the numerical analysis we use simulations of the signal passing
along a single line of sight directed at a galaxy cluster with a plasma
temperature of 7.5 keV. The signal in a single pixel consists of discrete values
$S_j$, $j=1,..,J$ in $J=384$
frequency channels $\nu_j$ covering the range from 10 GHz to 2.9 THz with a
channel width of $\Delta\nu=7.5$ GHz.

We assume that the angular resolution of the experiment, at least for nearby clusters and relatively high-frequency bands $\gtrsim 300$ GHz, is high enough for the signal source to be continuous.
Galaxy clusters contain most of their SZ signal in the brightest
central $\sim200$ kpc part.
This scale has apparent size larger than $\sim20$ arcsec at any redshift due to the
cosmological expansion.
With larger beam size the instrument can collect more foregrounds while collecting
almost the same amount of signal,
and this will decrease the signal to noise ratio. Smaller beams are technically more challenging,
since they require larger apertures. For a cluster
at 300 Mpc the central 200 kpc has angular size of 2 arcmin. A space observatory such as \cite{2021PhyU...64..386N} with a primary mirror size of 10 m has a diffraction limit of
$\sim 1$ arcmin at 100 GHz. 
Therefore below we give the required sensitivity in the units of Jy/sr assuming the source is
resolved.  This can be converted into Jy by assuming a pixel size equal to the apparent size of a central part of a cluster, i.e. $20$ arcsec for a distant cluster and $\sim2$ arcmin for a nearby cluster.

In accordance with the definitions and descriptions
of foregrounds in Secs. II and III, we consider eight components
whose spectral shapes are well defined ($M=8$),
three components with poorly defined spectra
($K=3$) and photon noise.
The list of components looks as follows:

{\it The well-defined spectra} ($f^m(\nu_j)$):\\
\hspace{-0.4cm}$m=1$: anisotropic SZ effect;\\
$m=2$: thermal SZ effect;\\
$m=3$,4,5,6: first four relativistic corrections to the thermal SZ
effect respectively;\\
$m=7$: spectrum of free-free transitions;\\
$m=8$: spectrum of the CMB anisotropy.

Here we do not take into account the spectrum of the CMB monopole,
since it can be easily removed from the observational
data.

{\it Poorly defined spectra} ($\tilde{f}^k(\nu_j,{\bf P_k})$):\\
\hspace{-0.4cm}$k=1$: dust emission with 2 free parameters $\boldsymbol{P_1}=(T_d,b_d)$, $\boldsymbol{P_1}\in\tilde{\Omega}_1$, see Eq. (7);\\
$k=2$: cosmic infrared background (CIB) with 2 free parameters
$\boldsymbol{P_2}=(T_c,b_c)$, $\boldsymbol{P_2}\in\tilde{\Omega}_2$, see Eq. (7);\\
$k=3$: synchrotron radiation with 1 free parameter $\boldsymbol{P_3}=b_s$,
 $\boldsymbol{P_3}\in\tilde{\Omega}_3$, see Eq. (8).

Here the possible ranges of parameters variation
$\tilde{\Omega}_1,\tilde{\Omega}_2,\tilde{\Omega}_3$
are defined in Section II (see also Fig. 1).

{\it Photon noise}:\\
We model photon noise as white noise with a diagonal covariance
matrix ${\bf C}=\langle{\bf N^T}{\bf N}\rangle$: $C_{ij}=\delta_i^j\sigma^2$.
Here $\sigma$ is the noise level per single channel. This value can
be considered as the sensitivity of the experiment.

In Fig. 2, we demonstrate the results of our numerical experiment using the
LRM method, previously tested in \citep{2024PhRvD.109b3523M}, for signal component separation.
We do not try to determine the contributions to the overall signal from poorly
defined components, i.e. from the CIB, dust, and synchrotron, owing to the lack
of information about these spectra.

The left panel shows the result of
estimating the signal component amplitudes $\alpha_m$ in the case when the
photon noise is 1 Jy/sr in each of the 384 channels. At this sensitivity, all considered components
with well-defined spectral shapes are recovered with good accuracy. The only
exception is the aSZ, for which the signal-to-noise ratio in this case is $\sim 1$.

The right panel shows the results of applying the LRM method to extract the aSZ signal depending on sensitivity, i.e. on $\sigma$.
We are interested in the sensitivity needed to find the signal of interest,
i.e. in this case the anisotropic SZ effect ($m=1$).
The weights $\boldsymbol{\omega^1}$ 
are always normalized in such a way that the response to the dipole part of the aSZ is equal to 1. This means that the total response to all other signal components, including noise, can be interpreted as the noise-to-signal ratio. For comparison, we also present the results of applying the cILC method, analogously to how it was done in
\citep{2024PhRvD.109b3523M}, where the extraction of the $\mu$ distortion from
the observed spectrum was considered. This approach gives a result that is approximately 1 order of magnitude worse. Furthermore, similar to the results demonstrated in \citep{2024PhRvD.109b3523M}, as the photon noise becomes less than $\sim 0.02$ Jy/sr, the cILC method reaches an asymptotic constant and ceases to improve the result. This occurs due to an insufficiency of derivatives with respect to parameters for
foregrounds with poorly defined spectra, which leads to an irreducible response to these components.  An increase in the number of derivatives with respect to parameters slightly improves the result for small photon noise and significantly degrades it if the photon noise is large.

Thus, when processing data using the LRM approach, a sensitivity of $1$ Jy/sr is
sufficient for the response to the aSZ signal, caused by the dipole anisotropy component, to be
equal to the response to foregrounds and noise.
For the quadrupole anisotropy, this is achieved with a sensitivity of $\sim$ 0.05 Jy/sr and for the octupole, with $\sigma\sim 0.01$ Jy/sr.

\section{Conclusions}

In our paper, we assessed the practical prospects for independent measurement
of the CMB anisotropy multipoles with $\ell=1,2,3$ using the anisotropic
Sunyaev-Zel'dovich effect.

The LRM method can
be effectively applied to any observational data containing components with
poorly defined spectra. Beyond studies of the aSZ, one of the most promising applications of this approach is the detection of $\mu$ spectral distortions in the CMB.

We have not yet taken into account the kinematic SZ effect.
Consequently, our estimate of the sensitivity required for aSZ detection
($\sim 0.05, 0.1$ Jy/sr) should be regarded as preliminary and will require further refinement. Adding the kinematic SZ effect and
its relativistic corrections will
increase the number of components that need to be separated
from each other. However, it not only enables detection of CMB anisotropy
at the scattering point, but also provides critical constraints on (1)
the cluster plasma temperature, (2) the cluster's peculiar velocity
relative to the CMB rest frame, and (3) its line-of-sight velocity component.

It is important to note that linear polarization occurs as a result of
scattering of anisotropic radiation by clusters. The
polarization produced as a result of ``cold" Thomson scattering in the nonrelativistic regime was considered in \citep{1997PhRvD..56.4511K} for
clusters at high redshifts to measure the CMB quadrupole, thereby getting
around cosmic variance. Nevertheless, the relic radiation itself
is also initially polarized starting from the recombination epoch. Thus,
without considering spectral distortions it is quite difficult to
distinguish between the polarization resulting from scattering on the SZ cluster
and the initial CMB polarization. In  \citep{2020PhRvD.101l3510N}, it is shown
that as a result of the aSZ, in addition to the distortion of
the spectral radiance $I(\nu)$ , a spectral distortion of the Stokes
parameters $Q(\nu)$ and $U(\nu)$ arises as well.
Unlike the spectral radiance, the spectrum of the Stokes parameters is
distorted only by the quadrupole and octupole of the anisotropy.
The dipole and, what is especially important, the CMB monopole do not
introduce additional terms. This means that the relativistic thermal
corrections from the dominant monopole do not contribute to the
distortions of
the polarization spectrum  and the task of component separation is
significantly simplified.

Future work should incorporate the kinematic Sunyaev-Zel'dovich effect, and it would be highly desirable to exploit the linear polarization of radiation from galaxy clusters.

\section*{ACKNOWLEDGMENTS}
We wish to thank the referee for valuable comments and productive discussion.
\section*{DATA AVAILABILITY}
There are no publicly available research data or software supporting this manuscript. Requests for further information or data should be sent to the authors.

\def\apj{Astrophys.~J}
\def\apjl{Astrophys.~J.,~Lett}
\def\apjs{Astrophys.~J.,~Supplement}
\def\an{Astron.~Nachr}
\def\aap{Astron.~Astrophys}
\def\mnras{Mon.~Not.~R.~Astron.~Soc}
\def\pasp{Publ.~Astron.~Soc.~Pac}
\def\aaps{Astron.~and Astrophys.,~Suppl.~Ser}
\def\apss{Astrophys.~Space.~Sci}
\def\ibvs{Inf.~Bull.~Variable~Stars}
\def\japa{J.~Astrophys.~Astron}
\def\na{New~Astron}
\def\aspproc{Proc.~ASP~conf.~ser.}
\def\aspcs{ASP~Conf.~Ser}
\def\aj{Astron.~J}
\def\actaa{Acta Astron}
\def\araa{Ann.~Rev.~Astron.~Astrophys}
\def\caosp{Contrib.~Astron.~Obs.~Skalnat{\'e}~Pleso}
\def\pasj{Publ.~Astron.~Soc.~Jpn}
\def\memsai{Mem.~Soc.~Astron.~Ital}
\def\astl{Astron.~Letters}
\def\aipproc{Proc.~AIP~conf.~ser.}
\def\physrep{Physics Reports}
\def\jcap{Journal of Cosmology and Astroparticle Physics}

\bibliography{a18asz.bib}



\end{document}